\documentclass{emulateapj}

\def\Mesz{M\'esz\'aros~}

\def\cm3{\mbox{cm}^{-3}}

\def\tdec2{t_{dec,2}}
\def\cm{\mbox{cm}}

\slugcomment{ }

\begin{document}
\title{Early soft X-ray to UV emission from double neutron star mergers: implications from the long-term radio and x-ray emissions of GW 170817 }

\author{Xiang-Yu Wang\altaffilmark{1,2}, Zhi-Qiu Huang\altaffilmark{1,2}}

\altaffiltext{1}{School of Astronomy and Space Science, Nanjing University, Nanjing 210093, China;
xywang@nju.edu.cn}
\altaffiltext{2}{Key laboratory of Modern Astronomy and Astrophysics (Nanjing University), Ministry of Education, Nanjing 210093, China}

\begin{abstract}
Recent long-term radio  follow-up observations of GW 170817 reveals a simple power-law
rising light curve, with a slope of $t^{0.78}$, up to 93 days after the merger. The latest X-ray detection at 109 days is also consistent with such a temporal slope. Such a shallow rise behavior requires a mildly relativistic outflow with a steep velocity gradient profile, so that slower material with larger energy catches up with the decelerating ejecta and re-energizes it.  It has been suggested that this mildly relativistic outflow may represent a cocoon of material. We  suggest that the velocity gradient profile may form during the stage that the cocoon  is breaking out of the  merger ejecta, resulted from shock propagation down a density gradient. The cooling of the hot relativistic cocoon material immediately after it breaks out  should have produced soft X-ray to UV radiation at  tens of seconds to hours after the merger. The soft X-ray emission has a luminosity of $L_{\rm X}\sim 10^{45}{\rm erg s^{-1}}$ over a period of tens of seconds for a merger event like GW 170817. The UV emission shows a rise initially and peaks at  about a few hours  with a luminosity of  $L_{\rm UV}\sim 10^{42} {\rm erg s^{-1}}$. The soft X-ray transients could be detected by future wide-angle X-ray detectors, such as the Chinese mission Einstein Probe. This soft X-ray/UV emission would serve as one of the earliest electromagnetic counterparts of gravitation waves from double neutron star mergers and could provide the earliest localization of the sources.

\end{abstract}


\section{Introduction}
The recent detection of gravitation waves (GW) from a  double neutron star (DNS) merger, known as GW 170817 \citep{Abbott2017a}, and the following detection
of an electromagnetic counterpart marks a new era for studying DNS mergers \citep{Abbott2017b}. The $\gamma$-ray satellite {\em Fermi} detected a sub-energetic short gamma-ray burst (GRB 170817A) about 2 seconds after the GW event \citep{Goldstein2017}. A macronova/kilonova was detected in the IR to UV bands since about 10 hours after the merger (e.g. \citep{Coulter2017,Drout2017,Evans2017}). Swift began searching for a counterpart to GW 170817
with its X-ray Telescope (XRT) from about 0.04 days after the merger (Evans et al. 2017), but  no new X-ray source was found. The {\it Chandra} X-ray satellite detected   X-ray counterparts at about 9  and 15.1 days after the merger \citep{Troja2017a,Margutti2017a}. Later on, radio emission was detected by VLA at about 16, 19 and 39 days after the merger \citep{Hallinan2017,Alexander2017}. The X-ray and radio emissions are thought to arise from the synchrotron emission of an expanding blast wave, which is powered by some ejecta from the DNS merger (e.g., \citet{Kasliwal2017,Murguia-Berthier2017}).

Very recently, long-term radio observations find that the radio flux continue to rise with a simple power-law in time, $F_\nu \propto t^{0.78\pm0.05}$ \citep{Mooley2017}. The spectrum is consistent with optically-thin synchrotron emission, with $F_\nu \propto \nu^{-0.61\pm0.05}$. The latest X-ray detection with {\it Chandra} of the source find that the X-ray emission is also brightening \citep{Troja2017b,Margutti2017b,Haggard2017}.
The X-ray brightening suggests a temporal slope  consistent  with the radio light curve.  Such a temporal slope can not be produced by a single-velocity ejecta, which would produce
a $t^3$ rising light curve or a decreasing light curve {for the observe frequency $\nu$ locating between the two break frequencies (i.e., $\nu_m<\nu<\nu_c$, \citet{Sari1998})},  as the ejecta is coasting or decelerating in {a constant density ISM} (e.g. \citet{Xiao2017}).  Depending on the density of the circum-burst medium, two velocity profiles of the outflow have been suggested to fit the radio data \citep{Mooley2017}. For a density of $n=0.03 {\rm cm^{-3}}$, the distribution of the kinetic energy is $E_k(>\gamma\beta)=5\times10^{50}{\rm erg} (\gamma\beta/0.4)^{-5}$, with a maximum velocity of $\beta_M=0.8$. While for a lower density of $n=10^{-4}{\rm cm^{-3}}$, $E_k(>\gamma\beta)=2\times10^{51}{\rm erg} (\gamma\beta)^{-5}$ with a maximum Lorentz factor of $\gamma_M=3.5$. Such velocity profile suggests that  the energy in the blast wave is increasing with time, which may be due to that  slower material with larger energy catches up with the decelerating ejecta and re-energizes it ({see also \citet{Pooley2017}}). To explain the emission of GRB 170817A  in the meantime, a relativistic outflow with $\gamma>2-3$ may be necessary \citep{Gottlieb2017}, so the low-density case is favored. The outflow then may represent the cocoon material \citep{Gottlieb2017}, which forms as the jet is propagating through the DNS merger ejecta \citep{Nagakura2014,Nakar&Piran2017,Lazzati2017a,Lazzati2017b}

This velocity profile may form at the stage when the cocoon breaks out of the merger ejecta. This profile reflects the structure of the outflow immediately after it break out, since after that the outflow matter simply undergoes free expansion. Such a self-similar velocity gradient profile is  common in core collapse supernovae. The supernova shock experiences acceleration in the steep density gradients of the progenitor envelope and the velocity gradient profile forms   after the shock breaks out of the  envelope \citep{Matzner&McKee1999}. We suggest that the velocity gradient profile in the case of GW 170817 forms when the cocoon breaks out of the DNS merger ejecta, resulted from shock propagation down a density gradient.

The cocoon is hot as it consists of mainly shock heated material from the merger ejecta. After the DNS merger ejecta has been shocked, its thermal and kinetic energies are approximately equal. The next phase of evolution is  postshock acceleration, in which heat is converted into outward motion and the cocoon material approaches a state of free expansion.  The internal energy will be released once the cocoon expands to a radius where it becomes transparent to radiation \citep{Nakar&Piran2017}. The physics is similar to that of a cooling envelope after the supernova shock breaks out. Recently, \citet{Piro2017}  show that  early $\la 4$ day optical/infrared emission of GW 170817 can be explained by shock cooling emission of the non-relativistic merger ejecta. \citet{Kisaka2015} shows that the early macronova/kilonova emission of GRB 130603B could be due to  shock cooling emission powered by  a central engine.  In this paper, we calculate the cooling emission from the mildly-relativistic shocked materials (i.e. the cocoon), taking into account the velocity gradient profile of the expanding materials. Due to that the cocoon material has a much higher velocity, compared with the bulk ejecta of the DNS merger, it becomes transparent much earlier and thus the  cooling emission constitutes the earliest electromagnetic counterpart of the DNS merger (only after the prompt $\gamma$-ray burst emission). In \S 2,  we discuss the constraints on the cocoon properties placed by recent long-term radio and X-ray observations. Then in \S 3, we calculate the light curves of the cocoon cooling emission and study the detectability by future wide-angle X-ray and UV missions. Finally we give discussions and conclusions.

\section{Constraining the cocoon properties with long-term radio and X-ray observations}

The radio follow-up observations of GW170817 reveals a steady rise in the light curve with a slope $F_\nu \propto t^{0.78\pm0.05}$ \citep{Mooley2017}.  Following the "refreshed" shock scenario of GRB afterglows \citep{Rees1998,Sari2000}, we study the velocity gradient of the outflow with this radio light curve. We assume that the source ejects  shells of a range of Lorentz factors,  with a mass profile of $m(>\gamma)\propto \gamma^{-s}$ in the range of $\gamma_m<\gamma<\gamma_M$, where $\gamma_m$ and $\gamma_M$ are, respectively, the minimum and maximum Lorentz factors of the ejected shells. For an observational frequency locating between the injection break frequency and the cooling frequency, i.e., $\nu_m<\nu<\nu_c$\citep{Sari1998}, the light curve of {the forward shock emission} should be \citep{Sari2000}
\begin{equation}
F_\nu\propto t^{-\frac{6-6s+24\delta}{2(7+s)}},
\end{equation}
where $\delta$ is the spectral index ($F_\nu\propto\nu^{-\delta}$). With an observed temporal slope of $F_\nu\propto t^{0.78}$ and a spectral slope of $\delta=0.61\pm0.05$, we obtain  $s\simeq7\pm0.5$. This value is in roughly agreement with $s=6$ obtained by \citet{Mooley2017}, who use numerical codes to calculate the light curve.  The radio flux can be used to place constraints
on the energy of the ejecta. The flux at 3 GHz is $F_\nu=151\pm39{\rm \mu {\rm Jy}}$ at $t=17.39$ days after the merger \citep{Hallinan2017}. We then obtain the kinetic energy of the blast wave
\begin{equation}
E_k\simeq 10^{49}{\rm erg} \epsilon_{e,-1}^{-0.92} \epsilon_{B,-2}^{-0.62} n_{-4}^{-0.38}(\frac{F_{\nu}}{151\mu {\rm Jy}})^{0.77}(\frac{t}{17.39{\rm d}})^{\frac{3s-3}{7+s}},
\end{equation}
where $\epsilon_e$ and $\epsilon_B$ are the electron energy and magnetic energy equipartition factors in the shock, and $n$ is the number density of the surrounding medium. Here we use the notations $\epsilon_{e,-1}=\epsilon_e/10^{-1}$, $\epsilon_{B,-2}=\epsilon_B/10^{-2}$ and $n_{-4}=n/(10^{-4} {\rm cm^{-3}})$.

{\it Chandra} re-observed the source at about 109 days after the merger \citep{Troja2017b,Margutti2017b,Haggard2017}.
The X-ray flux indicates a spectral slope $\delta \simeq0.6$ from radio to X-ray, which is consistent with the radio spectral slope \citep{Margutti2017a}. The spectrum of the X-ray emission alone is found to be $\delta =0.62\pm0.27$, in agreement with the global spectrum from radio to X-ray. This indicates that the cooling break frequency $\nu_c$ is above the X-ray band at $t=109 {\rm d}$. Using the condition $\nu_c ( t=109 {\rm d})>10^{18}{\rm Hz}$ and the energy in the blast wave given by Eq.(2), we get
\begin{equation}
n<1.6\times10^{-3} {\rm cm^{-3}}\epsilon_{e,-1}^{0.57}\epsilon_{B,-2}^{1.46}.
\end{equation}
{A low density of $n<0.04 {\rm cm^{-3}}$ for the surrounding medium has also been inferred from the limit on the mass of neutral hydrogen  \citep{Hallinan2017}.} {Note that the inverse Compton (IC) cooling  is not taken into account in our estimate of $\nu_c$. IC cooling can become important if $\epsilon_e\gg\epsilon_B$ and depends on the ratio between $\nu_m$ and $\nu_c$. Since the observations of GW170817 suggest that $\nu_m$ is below the radio band and $\nu_c$ is above the X-ray, the Compton parameter is $Y<1$ for reasonable values of $\epsilon_e$ and $\epsilon_B$ \citep{Wang2010,Beniamini2015}. Thus, the IC cooling does not introduce a significant change to our equation 3.}
Then we find a lower limit on the blast wave kinetic energy at  $t={\rm 17.39 {\rm d}}$,
\begin{equation}
E_k>3.5\times10^{48}{\rm erg} \epsilon_{e,-1}^{-1.1}\epsilon_{B,-2}^{-0.38}.
\end{equation}

The single power-law temporal behavior since $t=17.39$ days after the merger requires that the fastest shell has been decelerated before this time. From this, we obtain
\begin{equation}
\gamma_M\ga 3 E_{b,49}^{1/8} n_{-4}^{-1/8} (\frac{t}{17.39{\rm d}})^{-3/8}.
\end{equation}

{In the above, we have assumed that the forward shock dominates the flux from the possible reverse shock in the  refreshed shock scenario. This assumption is valid in our case since the injection break frequency $\nu_m$ of the reverse shock emission is much lower than that of the forward shock and the observe frequency is above $\nu_m$ of the forward shock emission  \citep{Sari2000}. \citet{Kumar&Piran2000} have shown that the Lorentz factor of the inner shell with respect to decelerating outer shell is about 1.25 when the collision takes place\footnote{The values of 1.25 corresponds to a the case that the outer, decelerating shell has a constant energy. For the case of re-energized, decelerating outer shell, the relative Lorentz factor is even lower (see Eq.3 in \citet{Rees1998}).}. So the Lorentz factor of the reverse shock is expected to be $\Gamma_{rs}< 1.25$. Then we expect that the ratio of thermal Lorentz factors of electrons in the reverse shock and forward shock is $\gamma_{e,rs}/\gamma_{e,fs}= (\Gamma_{rs}-1)/(\Gamma_{fs}-1)\la 0.1$, where $\Gamma_{fs}$ is the Lorentz factor of the forward shock. Although the number of electrons in the reverse shock is larger than that of the forward shock by a factor of $\Gamma_{fs}$ \citep{Sari2000}, the synchrotron flux of the forward shock at the frequency $\nu_m<\nu<\nu_c$ dominates over that of the reverse shock.}

\citet{Mooley2017} obtain  $E(>\gamma\beta)=2\times10^{51}(\gamma\beta)^{-5}$ with  a maximum Lorentz factor of $\gamma_M=3.5$ by fitting the radio data. Our simple analytic result is well consistent with their result. In the following calculations, we assume that the fastest shell has a Lorentz factor  of $\gamma_M=3.5$ and an energy of $E_M=4\times10^{48}{\rm erg}$, and assume that the shells has a velocity gradient profile of
$m(>\gamma\beta)\propto (\gamma\beta)^{-s}$ in the velocity range of $1\le \gamma\beta\le 3.5$.

\section{Soft X-ray and UV emission from the cooling cocoon}
This velocity profile reflects the structure of the cocoon material immediately after it breaks out of the  merger ejecta, since after that the cocoon material simply undergoes free expansion. When the cocoon material becomes transparent, the cooling of the cocoon material will produce an electromagnetic signal.

Assuming a cocoon has a self-similar mass distribution $m(>\gamma\beta)= m_M(\frac{\gamma\beta}{\gamma_M})^{-s}$. For $E_M=4\times10^{48}{\rm erg}$ and $\gamma_M=3.5$, the mass of the maximum velocity ejecta is $m_M=1.2\times10^{27} {\rm g}$. The highest velocity shell in the outmost region  becomes transparent at the earliest time. The shell with mass $m$ becomes transparent when its optical depth satisfies the condition
\begin{equation}
\tau(m)=\frac{\kappa m}{4\pi r^2}=\frac{c}{v_{sh}}\simeq 1 .
\end{equation}
With $r=2(\gamma \beta)^2 c t$, we can obtain the mass and Lorentz factor of the shell that  becomes transparent at a given time $t$, i.e.,
\begin{equation}
m=m_M (\frac{t}{t_M})^{\frac{2s}{4+s}}
\end{equation}
and
\begin{equation}
\gamma\beta=\gamma_M(\frac{t}{t_M})^{-\frac{2}{4+s}},
\end{equation}
where
\begin{equation}
t_M=(\frac{\kappa m_M}{16\pi \gamma_M^4 c^2})^{1/2}=12 {\rm s} E_{M,48.6}^{1/2} \gamma_{M,3.5}^{-2.5} \kappa_1^{1/2}
\end{equation}
is the characteristic timescale, corresponding the time when the fastest shell becomes transparent. Here we use the notations that $\kappa_1=\kappa/{\rm 1.0 cm^{2} g^{-1}}$, $E_{M,48.6}=E_M/10^{48.6}{\rm erg}$ and $\gamma_{M,3.5}=\gamma_M/3.5$.

The initial internal energy within each shell is roughly half of the final kinetic energy of the shell for mildly-relativistic shocks \citep{Tan2001}, so we assume $E_0(\gamma)={\frac{1}{2}}\gamma m c^2$. The internal energy $E$ decreases due to adiabatic expansion. Since $E=\gamma \varepsilon' V'$ and the comoving volume  is $V'\propto r^2 r/\gamma$, so $E\propto \varepsilon' r^3$, where $\varepsilon'$ is the comoving energy density. As the energy density scales as $\varepsilon'\propto n'^{4/3}\propto r^{-4}$,  where $n¡ä$ is the comoving baryon number density, we have  $E\propto r^{-1}$. Thus at a given time $t$, the internal energy is
\begin{equation}
E(t)=E_0 \frac{R}{r}=\frac{1}{2\gamma\beta} \frac{mcR}{t},
\end{equation}
where $R$ is the initial radius of the cocoon at which it breaks out of the merger ejecta. The value of $R$ is not well-known.  {A breakout radius of $\sim10^{10}-10^{11} {\rm cm}$ is estimated from the duration and lag of the prompt GRB \citep{Gottlieb2017}}. \citet{Kasliwal2017} find a breakout radius of $\sim 3\times10^{11} {\rm cm}$ in their numerical simulations of the DNS merger.
The bolometric luminosity is roughly
\begin{equation}
\begin{array}{ll}
L=\frac{E(t)}{t}=\frac{m_M c R}{2\gamma_M t_M^2} (\frac{t}{t_M})^{-\frac{6}{4+s}}\\
=3\times10^{45} {\rm erg s^{-1}} R_{11}
\kappa_1^{-1}\gamma_{M,3.5}^{3} (\frac{t}{t_M})^{-\frac{6}{4+s}}.
\end{array}
\end{equation}
The photosphere radius is
\begin{equation}
r_{ph}=(\frac{\kappa m}{4\pi})^{1/2}=9\times10^{12}{\rm cm} \kappa_{1}^{1/2}E_{M,48.6}^{1/2} \gamma_{M,3.5}^{-1/2}(\frac{t}{t_M})^{\frac{s}{4+s}}.
\end{equation}
Then, we obtain the effective photosphere temperature
\begin{equation}
\begin{array}{ll}
T_{\rm eff}=(\frac{\gamma^2 L}{4\pi r_{ph}^2 \sigma})^{1/4} \\=10^6 {\rm K} \,R_{11}^{1/4}\gamma_{M,3.5}^{3/2}\kappa_{1}^{-1/2}E_{M,48.6}^{-1/4} (\frac{t}{t_M})^{-\frac{1}{2}-\frac{1}{8+2s}},
\end{array}
\end{equation}
where $\sigma$ is the Stephan-Boltzmann constant.
Assuming a blackbody spectrum for the cooling emission, the observed luminosity at a given frequency $\nu$ is
\begin{equation}
L_\nu=\frac{8\pi^2 h \nu^3}{c^2}\frac{1}{exp(\frac{h\nu}{kT_{\rm eff}})-1}(\frac{r_{ph}}{D})^2
\end{equation}
where $D=1/(\gamma\beta)$ is the Doppler factor \citep{Gao2017}.
The cooling emission of relativistic matter  lasts until the  low velocity shell with $\gamma \beta=1$ becomes transparent, which is
\begin{equation}
t_{\rm rel}=t_M (\frac{\gamma \beta}{\gamma_M})^{-\frac{4+s}{2}}=6\times10^3 {\rm s} E_{M,48.6}^{-1/2}\gamma_{M,3.5}^{5/2}\kappa_{1}^{1/2}({\gamma\beta})^{-5},
\end{equation}
where we have taken $s=6$ in the last step. Note that the lowest value of $\gamma\beta$ is not known, as the radio and X-ray flux are still rising up to the present. However, $\gamma\beta$ should be $\ga1$, otherwise the  energy in cocoon would be too large.   After the time $t_{\rm rel}$, macronova/kilonova emission from sub-relativistic material of the bulk merger ejecta becomes dominated.  The bulk merger ejecta may also have some velocity gradient profile, as has been suggested in \citet{Piro2017} and\citet{Waxman2017}.

\begin{figure}
\centering
\includegraphics[scale=0.32]{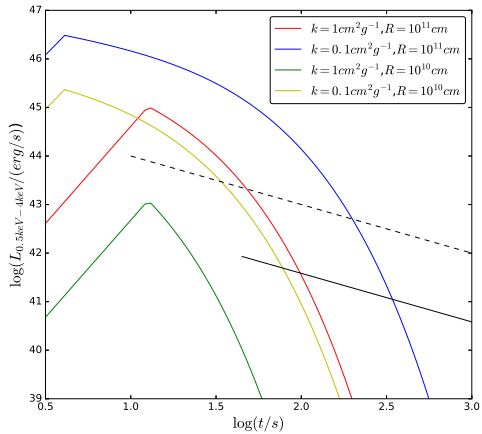}

\caption{Light curves of soft X-ray emission in 0.5-4 keV from the cooling of a cocoon, which has a velocity profile of $E_k(>\gamma\beta)=2\times10^{51}{\rm erg} (\gamma\beta)^{-5}$ with a maximum Lorentz factor of $\gamma_M=3.5$. The black dashed line represents the sensitivity curve of Einstein Probe telescope for a source at the same distance of GW 170817 (i.e., $d=40{\rm Mpc}$). The black solid straight line represents the  sensitivity curve of Swift XRT for a source at $d=40{\rm Mpc}$. The start time of XRT observation is taken to be 45 s (the average slew time for XRT) after a possible BAT trigger (note, however, that  GRB 170817A did not trigger Swift/BAT).  }
\end{figure}

\begin{figure}
\centering
\includegraphics[scale=0.32]{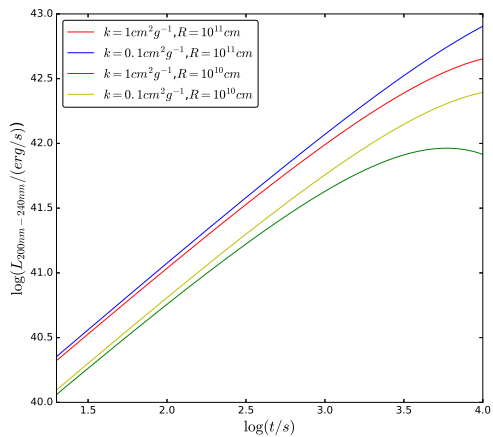}
\caption{Light curves of the   UV emission in 200-240nm from the cooling of the cocoon. The parameter values of the cocoon used in the calculation are the same as that used in Figure 1. Note that the cocoon emission lasts only until the lowest velocity shell with $\gamma\beta>1$ becomes transparent. See Eq. 15 in the text for the estimate of this time.}.
\end{figure}

Using the above formulas, we calculate the light curves of the soft X-ray emission in the energy range of $0.5-4$ KeV
and  UV light curves in 200-240nm.  The energy ranges are selected according to the future wide-angle X-ray and UV missions respectively, such as
Einstein Probe (EP) and ULTRASAT \citep{Yuan2015,Sagiv2014}.  The mostly poorly-known parameters in the calculation are the opacity $\kappa$ and the break-out radius $R$, thus we consider various combinations of these two parameters. The light curves of the soft X-ray emission are shown in Figure 1. To study the detectability of this emission, we also plot the sensitivity curve of EP telescope in the figure. For a wide range of the parameter space
of $\kappa$ and $R$, the soft X-ray emission lasts at least tens of seconds above the sensitivity of EP telescope if a similar event to GW 170817 occurs in the future. As this emission lags only a few seconds behind the merger, this would represent the earliest electromagnetic counterpart of  gravitation waves from DNS mergers (only after the prompt GRB signal). If Swift BAT is triggered by the prompt GRB (which is not the case for GRB 170817A), Swift X-ray  Telescope (XRT) could slew to the target in 20-70 s. The sensitivity curve of XRT is also shown in Figure 1, which indicates that the soft X-ray transient from cooling cocoon could be detected by XRT.

The UV light curves for the same parameter values are shown in Figure 2. The flux shows a monotonic rise in time because the peak frequency of the blackbody spectrum is above the UV frequency range at such early time. The luminosity in 200-240nm at the peak is above $10^{42}{\rm erg s^{-1}}$ even for the conservative parameter values.  The UV emission from the cocoon could be detected to distances above $\sim 1{\rm Gpc}$ by ULTRASAT, considering that the limit magnitude of ULTRASAT is $m=21$ \citep{Sagiv2014}. Such early UV emission could also be detected by Swift UVOT, if UVOT slews quickly enough to the target.  At later time, the UV emission will transit to the phase where the bulk ejecta of macronova/kilonova becomes dominated \citep{Metzger2017,Piro2017,Yu2017,Waxman2017}.

\section{Summary and Discussions }

We have suggested that soft X-ray to UV emission may be produced by  cocoons in DNS mergers at very early times after the merger, arising from the  cooling of the hot cocoon, similar to the cooling envelope emission in core collapse supernovae. As cocoons have  wide angles, these soft X-ray and UV transients also have wide angles and thus they have much better chance to be detected than the jet emission. Such soft X-ray and UV transients, if detected, would serve as  very early electromagnetic counterparts to the GW sources. As  X-ray and UV detectors have much better localization ability than the $\gamma$-ray detectors, they could provide accurate spatial position of GW events within minutes to hours. This will be important for further follow-up observations of the GW source.

We have assumed that the outflow powering the long-term radio and X-ray emission are relativistic with a maximum Lorentz factor of $\gamma_M\ga 3$, consistent with the cocoon scenario. It has been pointed out that a semi-relativistic outflow with a maximum velocity $\beta_{M}=0.8$ \citep{Mooley2017}, arising from the dynamic ejecta of the DNS merger,  can also explain the long-term radio and X-ray emission. In this case, we would not expect soft X-ray emission at early time, because the effective photosphere temperature  decreases as the maximum velocity decreases. However, we would still expect bright UV radiation from the cooling of this semi-relativistic ejecta on a timescale hours after the merger.

The prompt emission of GRB 170817A has a main pulse, followed by a weak and soft component {with a low signal to noise ratio} \citep{Goldstein2017,Zhang2017,Lu2017}. The soft component has  a luminosity of $L_{\gamma}=10^{46}{\rm erg s^{-1}}$ lasting for $\simeq 1.1$ s. The spectrum of this soft component can be fit by a blackbody with a temperature of $kT=10.3\pm1.5{\rm keV}$  \citep{Goldstein2017,Zhang2017}, had. It was suggested that this thermal component could be due to the photosphere emission from a cocoon (Goldstein et al. 2017). However, a straightforward calculation of the photosphere radius gives $r_{ph}=\gamma_c(L_\gamma/4\pi \sigma T^4)^{1/2}=3\times10^{8}(\gamma_c/3){\rm cm}$, which is too small for a  cocoon with a mildly-relativistic Lorentz factor of $\gamma_c=3$.

The spectrum of the soft X-ray emission in our case should be thermal  in the absence of dissipation below the photosphere, which can be  distinguished from non-thermal X-ray emission predicted in some models. For example, \citet{Zhang2013} suggests that if the post-merger product of DNS is a highly magnetized, rapidly rotating neutron  star, the dissipation of a proto-magnetar wind after the merger could produce non-thermal X-ray emission (see also \citet{Sun2017}).

\acknowledgments We thank Yunwei Yu, Zhuo Li and Bing Zhang for valuable discussions. This work is supported by the 973 program under grant 2014CB845800 and the NSFC under grant 11625312.


\begin{thebibliography}{99}
\bibitem[Abbott et al. (2017a)] {Abbott2017a} Abbott, B. P. et al., 2017, \prl, 119, 161101

\bibitem[Abbott et al. (2017b)] {Abbott2017b} Abbott, B. P. et al., 2017, ApJ, 848, L13

\bibitem[Alexander et al. (2017)]{Alexander2017} Alexander, K. D.; Berger, E.; Fong, W. et al., 2017, The Astrophysical Journal Letters, 848, L21

\bibitem [Beniamini et al. (2015)]{Beniamini2015}
Beniamini, P.; Nava, L.; Duran, R.; Piran, T., 2015, MNRAS, 454,1073

\bibitem[Coulter et al. (2017)] {Coulter2017} Coulter D. A., Foley, R. J., Kilpatrick, C. D., et al. 2017, Science, doi:10.1126/science.aap9811, (arXiv:1710.05452)


\bibitem[Drout et al. (2017)] {Drout2017}Drout, M. R., Piro, A. L., Shappee, B. J., et al. 2017, Science, doi:10.1126/science.aaq0049, (arXiv:1710.05443)


\bibitem[Evans et al. (2017)] {Evans2017}Evans, P., Cenko, S., Kennea, J. A., et al. 2017, Science, doi:10.1126/science.aap9580
(arXiv:1710.05437)

\bibitem[Gao et al. (2017)]{Gao2017} Gao, H., Cao, Z., Zhang, B., 2017,  arXiv:1711.08577

\bibitem[Goldstein et al. (2017)]{Goldstein2017} Goldstein, A., Veres P., Burns, E., et al. 2017, ApJL in press

\bibitem[Gottlieb et al. (2017)]{Gottlieb2017} Gottlieb, O., Nakar, E., Piran, T., Hotokezaka, 2017, arXiv:1710.05896


\bibitem [Haggard et al. (2017)]{Haggard2017} Haggard, D., et al., 2017,  GCN CIRCULAR  2220

\bibitem[Hallinan et al. (2017)] {Hallinan2017} Hallinan, G., Corsi, A., Mooley, K. P., et al. 2017, Science, doi:10.1126/science.aap9855 (arXiv:1710.05435)

\bibitem[Kasliwal et al. (2017)]{Kasliwal2017} Kasliwal, M. M., Nakar, E., Singer, L. P., et al. 2017, Science, doi:10.1126/science.aap9455 (arXiv:1710.05436)

\bibitem[Kisaka et al. (2015) ] {Kisaka2015}
Kisaka, S.; Ioka, K.; Takami, H., 2015, The Astrophysical Journal,  802, 119

\bibitem[Kumar \& Piran (2000)]{Kumar&Piran2000} Kumar, P., Piran, T., 2000, ApJ, 532, 286


\bibitem[Lazzati et al. (2017a)]{Lazzati2017a} Lazzati, D., Deich, A., et al., 2017a, MNRAS, 471, 1652

\bibitem[Lazzati et al. (2017b)]{Lazzati2017b} Lazzati, D., et al., 2017b, arXiv:1709.01468

\bibitem[Lu et al. (2017)] {Lu2017} Lu, R. J. et al., 2017, arXiv:1710.06979



\bibitem[Margutti et al. (2017a)]{Margutti2017a} Margutti, R.; Berger, E, et al., 2017a, The Astrophysical Journal Letters, 848,  L20

\bibitem[Margutti et al. (2017b)]{Margutti2017b} Margutti, R., et al. 2017b,   GCN CIRCULAR
22203

\bibitem [Matzner \& McKee (1999)]{Matzner&McKee1999} Matzner, C. D. \& McKee, C. F., 1999, \apj, 510, 379


\bibitem[Metzger et al. (2017)]{Metzger2017} Metzger, B. D., 2017, arXiv:1710.05931

\bibitem[Mooley et al. (2017)]{Mooley2017} Mooley, K. P., et al, 2017, arXiv:1711.11573

\bibitem[Murguia-Berthier(2017)]{Murguia-Berthier2017} Murguia-Berthier, A.; Ramirez-Ruiz, E., et al., 2017, The Astrophysical Journal Letters, 848, L34
\bibitem[Nagakura et al. (2014)]{Nagakura2014} Nagakura, H. et al., 2014, The Astrophysical Journal Letters,  784,   L28

\bibitem[Nakar \& Piran (2017)]{Nakar&Piran2017}Nakar, E., \& Piran, T., 2017 The Astrophysical Journal, 834,  28

\bibitem[Piro \& Kollmeier (2017)]{Piro2017} Piro, A. L., Kollmeier, J. A., 2017, arXiv:1710.05822

\bibitem[Pooley et al. (2017)]{Pooley2017} Pooley, D.; Kumar, P.; Wheeler, J. C., 2017, arXiv:1712.03240

\bibitem[Rees \& \Mesz (1998)]{Rees1998} Rees, M. J., \& \Mesz, P., 1998, ApJ, 496, L1

\bibitem[Sagiv et al. (2014)]{Sagiv2014}Sagiv, I., et al., AJ, 147, 79

\bibitem[Sari \& \Mesz (2000)]{Sari2000}Sari, R., \& \Mesz, P., 2000, The Astrophysical Journal,  535,  L33

\bibitem[Sari et al. (1998)]{Sari1998} Sari, R., Piran, T., Narayan, R. 1998, ApJL, 497, L17

\bibitem[Sun et al. (2017)]{Sun2017}Sun, H., et al., 2017, The Astrophysical Journal, 835, 7

\bibitem[Tan et al. (2001)]{Tan2001} Tan, J. C.; Matzner, . D.; McKee, C. F., 2001, The Astrophysical Journal,  551,  946

\bibitem[Troja et al. (2017a)] {Troja2017a} Troja, E., Piro, L., van Eerten, H. et al., 2017, Nature, 551, 71

\bibitem[Troja et al. (2017b)] {Troja2017b} Troja, E. et al. 2017b, GCN CIRCULAR 22201

\bibitem[Wang et al. (2010)]{Wang2010} Wang, X. Y., et al., 2010, ApJ, 712, 1232

\bibitem[Waxman et al. (2017)]{Waxman2017} Waxman, E., et al., 2017, arXiv:1711.09638

\bibitem[Xiao et al.(2017)]{Xiao2017} Xiao, D., Liu, L. D., Dai, Z. G., Wu, X. F., 2017, The Astrophysical Journal Letters,  850, L41

\bibitem[Yu \& Dai (2017)]{Yu2017}Yu, Y. W., Dai, Z. G., 2017, arXiv:1711.01898

\bibitem[Yuan et al. (2015)]{Yuan2015} Yuan, W, et al., 2015, arXiv:1506.07735


\bibitem[Zhang (2013)]{Zhang2013} Zhang, B., 2013, The Astrophysical Journal Letters, 763, L22

\bibitem[Zhang et al. (2017)]{Zhang2017} Zhang, B.-B., et al., 2017, arXiv:1710.05851
\end{thebibliography}
\end{document}